# DYNAMICAL CHARACTERISTICS OF ELECTROMAGNETIC FIELD UNDER CONDITIONS OF TOTAL REFLECTION

Aleksandr Ya. Bekshaev

Odessa I.I. Mechnikov National University
Dvorianska 2, 65082 Odessa, Ukraine
E-mail: bekshaev@onu.edu.ua

**Abstract**

The dynamical characteristics of electromagnetic fields include energy, momentum, angular momentum (spin) and helicity. We analyze their spatial distributions near the planar interface between two transparent and non-dispersive media, when the incident monochromatic plane wave with arbitrary polarization is totally reflected, and an evanescent wave is formed in the medium with lower optical density. Based on the recent arguments in favor of the Minkowski definition of the electromagnetic momentum in a material medium [*Phys. Rev. A* **83**, 013823 (2011); **86**, 055802 (2012); *Phys. Rev. Lett*. **119**, 073901 (2017)], we derive the explicit expressions for the dynamical characteristics in both media, with special attention to their behavior at the interface. Especially, the "extraordinary" spin and momentum components orthogonal to the plane of incidence are described, and the canonical (spin – orbital) momentum decomposition is performed that contains no singular terms. The field energy, helicity, the spin momentum and orbital momentum components are everywhere regular but experience discontinuities at the interface; the spin components parallel to the interface appear to be continuous, which testifies for the consistency of the adopted Minkowski picture. The results supply a meaningful example of the electromagnetic momentum decomposition, with separation of spatial and polarization degrees of freedom, in inhomogeneous media, and can be used in engineering the structured fields designed for optical sorting, dispatching and micromanipulation.

Keywords: total reflection, evanescent wave, polarization, electromagnetic momentum, spin-orbit decomposition.

**1. Introduction**

In the recent years, properties of structured optical fields have attracted growing attention of the scientific community. This attention is stimulated by fundamental considerations aimed at the deeper inspection and further refinement of general aspects of physical fields [1–10] as well as by more practical needs of optical imaging, probing and micromanipulation techniques [8–14]. In all such investigations, main efforts are directed to understanding the principles of formation and regulation of the field dynamical characteristics: energy, momentum and angular momentum (spin); frequently, in this context, the optical field chirality or helicity are considered which are closely related to the momentum and spin [12,15].

Among the great variety of physically meaningful and practically important structured fields, a significant place belongs to the evanescent waves (EW) that appear at the interface between two media with different electromagnetic properties (see, e.g., Refs. [16–22]). This relatively simple and ubiquitous field configuration enabled the comprehensive studies of basic extraordinary properties of structured fields – the transverse polarization-independent spin and the transverse polarization-dependent momentum orthogonal to the propagation direction [17,18,23]. Further analysis of the EW fields contributed to understanding the dynamical meaning of these extraordinary properties

[18–23]. Moreover, it stimulated experimental research [24] that revealed the direct mechanical manifestation of some of the field quantities previously considered as pure theoretical (virtual) constructions: the Belinfante (spin) momentum [18,25] and the reactive momentum (imaginary Poynting vector) [1,7,18,24].

Due to a series of recent works [17–23], the peculiar features of the spin and momentum distributions in the EW are well established; however, most of the analyses deal with the EW 'per se', regardless of the way the EW is formed by and how it is related with the electromagnetic field outside the EW region. This approach is reasonable in case of surface polariton waves [17,19–21], and it demonstrated some intriguing features, in particular, the spatial discontinuities and even singularities of the spin and momentum distributions at the interface [17]. Additional difficulties are associated with the difference between the Abraham's and Minkowski's definitions [26,27] of the field dynamical characteristics. As a rule, it entails no more than quantitative consequences for the usual 3D-volume (spatially unlimited) waves, but leads to striking discrepancies and re-builds the whole physical picture of the phenomena for the surface plasmon-polariton modes [19–21] (up to the sign reversal of the field momentum or spin). But more important is that the known analyses are not applicable to the most generic case of evanescent fields formed in the course of total reflection (TR) where the EW should be considered in conjunction with the non-decaying propagating waves in the optically denser medium – the situation that can be easily realized and is widely used for the EW investigation [16,24].

That is why in this paper we specially consider the field structure, energy flow pattern and the field dynamical characteristics for the situation where the plane wave is totally reflected at the interface between two dielectric media. The important step is to choose the correct expression for the field momentum and angular momentum in a material medium where the existence of controversial Abraham's and Minkowski's definitions [26,27] serves a source of additional indeterminacy and arbitrariness in any formal approach [20,22]. Here this is made on the base of the recently reported rigorous methodology of Philbin and Allanson who derived expressions for the electromagnetic field dynamical characteristics based on the field Lagrangian and the Noether's theorem [28–30] (their general results have been essentially supported and further developed with including of the canonical spin-orbit momentum decomposition and employing the microscopic analysis of the EWs in surface plasmon-polariton fields [31–33]). In this work, we show that in case of negligible dispersion of the contacting media, this methodology substantiates the validity of the Minkowski definition, which contradicts to the most of previous works where the dynamical characteristics were treated on the base of the Abraham definition or the Abraham-based and Minkowski-based quantities were involved on equal footing [19–21]. (Consequently, the Abraham-based analysis of the dynamical characteristics in the TR situation presented in [34] can be used for comparison and appropriate references but not for direct calculation of physical parameters). Having thus obtained the general definitions of the field dynamical characteristics in both media, explicit expressions are derived for the energy, helicity [18,30,31] and the momentum spatial distributions, and the momentum decomposition into the "spin" and "orbital" parts is consistently performed. The distributions obtained are analyzed with the special attention to the meaning and the nature of physical relations between the fields in different media.

## 2. Field configuration

In this section, we summarize the general expressions for the electric and magnetic fields in waves that participate in the TR process (see, e.g., [1,16]), and represent them in the unified form suitable for the further consideration. Let us consider a general model of the EW formation due to the TR of a monochromatic light wave with frequency $\omega$ where the electric and magnetic fields behave in time as $\text{Re}[\mathbf{E}\exp(-i\omega t)]$, $\text{Re}[\mathbf{H}\exp(-i\omega t)]$. Let the incident plane wave come from the lower half-space $z < 0$ (medium 1) with dielectric and magnetic constants $\varepsilon_1$, $\mu_1$ and refraction index $n_1 = \sqrt{\varepsilon_1 \mu_1}$, and the boundary $z = 0$ separates it from the medium 2 ($z > 0$) with parameters $\varepsilon_2$, $\mu_2$ and $n_2 = \sqrt{\varepsilon_2 \mu_2}$ (see figure 1). The wave direction is specified by the wave vector $\mathbf{k}_1 = k_1 (\sin\theta_1, 0, \cos\theta_1)^T$, $\theta_1$ is the angle of incidence, $k_1 = n_1(\omega/c)$ and $k_2 = n_2(\omega/c)$ are the wavenumbers in media 1 and 2, respectively, $c$ is the light velocity in vacuum.

1) The electric and magnetic vectors in the incident wave are described by equations

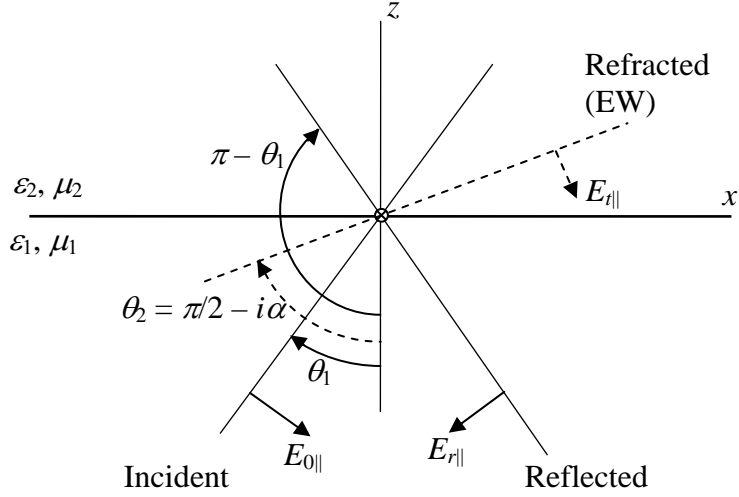

**Figure 1**. Geometrical configuration of the waves near the interface $z = 0$ and the coordinate frame (axis $y$ is orthogonal to the figure plane and directed along the view line). According to (5), the direction of the refracted (evanescent) wave is formally characterized by the complex angle, which is schematically shown by dashed lines.

$$\mathbf{E}_0 = \begin{pmatrix} E_{0\|} \cos\theta_1 \\ E_{0\perp} \\ -E_{0\|} \sin\theta_1 \end{pmatrix} e^{ik_1(z\cos\theta_1 + x\sin\theta_1)}, \quad \mathbf{H}_0 = \sqrt{\frac{\varepsilon_1}{\mu_1}} \begin{pmatrix} -E_{0\perp} \cos\theta_1 \\ E_{0\|} \\ E_{0\perp} \sin\theta_1 \end{pmatrix} e^{ik_1(z\cos\theta_1 + x\sin\theta_1)} \quad (1)$$

where $E_{0\|}$ and $E_{0\perp}$ are the electric strength components parallel ($p$) and orthogonal ($s$) to the plane of incidence ($p$- and $s$-polarized components).

2) Reflected wave differs by replacement $\theta_1 \to \pi - \theta_1$ and by the transformed field amplitudes:

$$\mathbf{E}_r = \begin{pmatrix} -E_{r\|} \cos\theta_1 \\ E_{r\perp} \\ -E_{r\|} \sin\theta_1 \end{pmatrix} e^{ik_1(-z\cos\theta_1 + x\sin\theta_1)}, \quad \mathbf{H}_r = \sqrt{\frac{\varepsilon_1}{\mu_1}} \begin{pmatrix} E_{r\perp} \cos\theta_1 \\ E_{r\|} \\ E_{r\perp} \sin\theta_1 \end{pmatrix} e^{ik_1(-z\cos\theta_1 + x\sin\theta_1)} \quad (2)$$

where

$$E_{r\perp} = R_\perp E_{0\perp}, \quad E_{r\|} = R_\| E_{0\|}, \quad (3)$$

and

$$R_\perp = \frac{\sqrt{\varepsilon_1/\mu_1} \cos\theta_1 - i\sqrt{\varepsilon_2/\mu_2} \sinh\alpha}{\sqrt{\varepsilon_1/\mu_1} \cos\theta_1 + i\sqrt{\varepsilon_2/\mu_2} \sinh\alpha}, \quad R_\| = \frac{\sqrt{\varepsilon_2/\mu_2} \cos\theta_1 - i\sqrt{\varepsilon_1/\mu_1} \sinh\alpha}{\sqrt{\varepsilon_2/\mu_2} \cos\theta_1 + i\sqrt{\varepsilon_1/\mu_1} \sinh\alpha} \quad (4)$$

are the reflection coefficients, $|R_\perp| = |R_\|| = 1$. Equations (4) follow from formal application of the well known Fresnel formulas and the Snell's law [1] to the TR case:

$$\sin\theta_2 = \frac{n_1}{n_2}\sin\theta_1 = \cosh\alpha > 1, \quad \cos\theta_2 = i\sqrt{\left(\frac{n_1}{n_2}\right)^2 \sin^2\theta_1 - 1} = i\sinh\alpha, \tag{5}$$

which corresponds to the complex refraction angle $\theta_2 = \frac{\pi}{2} - i\alpha$.

3) Similarly, the refracted (evanescent) wave is described by relations [1,35]

$$\mathbf{E}_t = \begin{pmatrix} iE_{t\|}\sinh\alpha \\ E_{t\perp} \\ -E_{t\|}\cosh\alpha \end{pmatrix} \exp(ik_2 x \cosh\alpha - k_2 z \sinh\alpha),$$

$$\mathbf{H}_t = \sqrt{\frac{\varepsilon_2}{\mu_2}} \begin{pmatrix} -iE_{t\perp}\sinh\alpha \\ E_{t\|} \\ E_{t\perp}\cosh\alpha \end{pmatrix} \exp(ik_2 x \cosh\alpha - k_2 z \sinh\alpha) \tag{6}$$

where

$$E_{t\perp} = T_\perp E_{0\perp}, \quad E_{t\|} = T_\| E_{0\|}, \tag{7}$$

and

$$T_\perp = \frac{2\sqrt{\varepsilon_1/\mu_1}\cos\theta_1}{\sqrt{\varepsilon_1/\mu_1}\cos\theta_1 + i\sqrt{\varepsilon_2/\mu_2}\sinh\alpha}, \quad T_\| = \frac{2\sqrt{\varepsilon_1/\mu_1}\cos\theta_1}{\sqrt{\varepsilon_2/\mu_2}\cos\theta_1 + i\sqrt{\varepsilon_1/\mu_1}\sinh\alpha}. \tag{8}$$

Noteworthy, the field (6) can be considered as a plane wave with wavevector $\mathbf{k}_2$ that lies in the (*XZ*) plane but possesses an imaginary *z*-component [1]; here it is represented in the form [35]

$$\mathbf{k}_2 = (k_2 \cosh\alpha, 0, ik_2 \sinh\alpha). \tag{9}$$

It is suitable to describe the polarization of the incident, reflected and evanescent waves via the polarization parameters [7,18]

$$m_0 = \frac{E_{0\perp}}{E_{0\|}}, \quad m_r = -\frac{R_\perp}{R_\|} m_0, \quad m_t = \frac{T_\perp}{T_\|} m_0 \tag{10}$$

and the associated Stokes parameters

$$\tau_0 = \frac{1 - |m_0|^2}{1 + |m_0|^2}, \quad \chi_0 = \frac{2\,\mathrm{Re}\,m_0}{1 + |m_0|^2}, \quad \sigma_0 = \frac{2\,\mathrm{Im}\,m_0}{1 + |m_0|^2} \tag{11}$$

(similar equations for the transmitted and reflected waves can be written including $m_t$, $\tau_t$, $\chi_t$, $\sigma_t$ and $m_r$, $\tau_r$, $\chi_r$, $\sigma_r$) that characterize the degree of *s–p*, ±45° and circular polarization, correspondingly.

Usually, relations (4), (6) and (8) are represented in a more compact form using the wavevector components of (9) but in this paper we keep the complex-angle representation that preserves the direct genetic links between the angle of incidence $\theta_1$ and the imaginary "angle" $i\alpha$.

## 3. Dynamical characteristics of the optical field in presence of the TR interface

### 3.1. General expressions

Now consider the dynamical characteristics of the electromagnetic field in both regions 1 and 2: spatial distributions of the energy, momentum and angular momentum (spin). We start with the energy density that is determined by the known relation [1] (here trivially modified to comply with the Gaussian system of units)

$$w = \frac{g}{2}\left(\varepsilon |\mathbf{E}|^2 + \mu |\mathbf{H}|^2\right) \tag{12}$$

where $g = (8\pi)^{-1}$. However, in all material media, where coefficients $\varepsilon$ and $\mu$ depend on frequency, the effective energy density is described by the transformed formula (12) in which the permittivity and permeability are replaced along the rules $\varepsilon \to \varepsilon + \omega(d\varepsilon/d\omega)$, $\mu \to \mu + \omega(d\mu/d\omega)$ [1]. The terms $\omega(d\varepsilon/d\omega)$, $\omega(d\mu/d\omega)$ are responsible for the material dispersion and, according to [28,29], the similar terms describe the dispersion-induced contributions to the field momentum and angular momentum. Their presence introduces additional material parameters associated with the great variety of possible dispersion laws and thus hampers the general analysis of the TR fields. To avoid this difficulty, we note that in many TR situations the contacting media show rather low dispersion. For transparent dielectrics, this is typical; for example, dispersion of optical glasses in the most usable visible range (vacuum wavelength $\lambda \sim 500$ nm) is $dn/d\lambda \sim -6 \cdot 10^2$ cm$^{-1}$ [36], and assuming $\varepsilon = n^2$, $n \approx 1.5$, the relative "weight" of the dispersion correction can be found as

$$\frac{\omega}{\varepsilon}\frac{d\varepsilon}{d\omega} = -\frac{2\lambda}{n}\frac{dn}{d\lambda} \approx 0.04.$$

This is not completely negligible but clearly testifies that in many cases the dispersion can but weakly modify the quantitative results. Therefore, in the present consideration we will discard the dispersion terms at all.

With this prerequisite, the field momentum expression obtained in [28] reduces to

$$\mathbf{p} = \frac{g}{c}\text{Re}\left(\varepsilon\mu\mathbf{E}^* \times \mathbf{H}\right), \tag{13}$$

which is just the Minkowski momentum [1,26,27]. Then, after the standard decomposition based on the Maxwell equations in dielectric media (without free charges and currents) [2,7–9,31], expression (13) can be transformed to the familiar sum of the spin and orbital parts,

$$\mathbf{p} = \mathbf{p}^S + \mathbf{p}^O, \tag{14}$$

where

$$\mathbf{p}^S = \frac{g}{4\omega}\nabla \times \text{Im}\left(\varepsilon\mathbf{E}^* \times \mathbf{E} + \mu\mathbf{H}^* \times \mathbf{H}\right) = \frac{1}{2}\nabla \times \mathbf{s} \tag{15}$$

is the spin momentum and

$$\mathbf{p}^O = \frac{g}{2\omega}\text{Im}\left[\varepsilon\mathbf{E}^* \cdot (\nabla)\mathbf{E} + \mu\mathbf{H}^* \cdot (\nabla)\mathbf{H}\right] \tag{16}$$

is the orbital (canonical) momentum of the field in the medium (the Berry's notation [2] is used, $\mathbf{E}^* \cdot (\nabla)\mathbf{E} = E_x^*\nabla E_x + E_y^*\nabla E_y + E_z^*\nabla E_z$). Note that the spin-orbital decomposition of the momentum (13) naturally leads to the spin momentum (15) associated with the Minkowski spin

$$\mathbf{s} = \frac{g}{2\omega}\text{Im}\left(\varepsilon\mathbf{E}^* \times \mathbf{E} + \mu\mathbf{H}^* \times \mathbf{H}\right). \tag{17}$$

rather than with the Abraham spin density used before [7,17,18,34].

Usual interpretation of the momentum decomposition (14) [9,18,37] is that the orbital momentum is the canonical momentum directly following from the Noether's theorem and associated with the energy transport in the field whereas the spin momentum appears due to the divergenceless internal currents emerging because of inhomogeneous elliptical polarization. The orbital momentum is observable in the dipole light–matter interactions and produces the usual light pressure; the spin momentum is "virtual" in the sense that its mechanical action can be detected only in higher multipole orders.

The kinetic aspect of the field momentum, associated with the field energy transfer, is expressed by the Abraham momentum [31,32], and the energy flow density is determined by the Poynting vector $gc\,\text{Re}\left(\mathbf{E}^* \times \mathbf{H}\right)$ which can also be decomposed into the spin and orbital parts similar to (15) and (16) [7,17,18]. Herewith, the divergence-free spin part does not contribute to the real energy transport and the local energy flow is characterized by the orbital Abraham momentum $\mathbf{p}_A^O$ [37] which in our conditions can be naturally associated with the canonical momentum (16):

$$\mathbf{q} = c^2 \mathbf{p}_A^O = \frac{c^2}{\varepsilon\mu} \mathbf{p}^O. \tag{18}$$

The angular momentum density follows from the results of Refs. [29,31,32], and in the considered case of negligible dispersion can be found directly from the formal definition

$$\mathbf{j} = \mathbf{r} \times \mathbf{p} = \mathbf{r} \times \mathbf{p}^S + \mathbf{r} \times \mathbf{p}^O. \tag{19}$$

Here, the first summand is transformed with allowance for (15); besides, we take into account that due to non-local relation

$$\int \mathbf{r} \times (\nabla \times \mathbf{s}) dV = 2 \int \mathbf{s} dV$$

(integration is performed over the whole space and all the involved functions and derivatives are supposed to properly vanish at the infinity), the density terms in the form $\mathbf{r} \times (\nabla \times \mathbf{s})$ can be replaced by $2\mathbf{s}$. As a result, the first summand of (19), indeed, expresses the spin angular momentum density in the medium, and the above reasoning together with the results of [28,29] validate the Minkowski spin (17) to be the correct spin density in the material medium.

The second summand of (19) represents the orbital momentum density

$$\mathbf{l} = \frac{g}{2\omega} \mathbf{r} \times \text{Im}\left[\varepsilon \mathbf{E}^* \cdot (\nabla) \mathbf{E} + \mu \mathbf{H}^* \cdot (\nabla) \mathbf{H}\right]. \tag{20}$$

The details of the orbital angular momentum distribution are of crucial importance for many structured fields [9,14], in particular, for the EWs associated with the surface polaritons [17,19–23,31–33]. However, in case of TR this quantity is not well-defined in view of existence of the non-vanishing field formed by the incident and reflected waves (1), (2) and filling the half-space $z < 0$ so that the integral of the density (20) over the whole space diverges. Therefore, in this paper the further analysis of expression (20) will not be performed, and it is presented here just for the completeness and suitability of further references.

In addition to the energy and momentum, an important dynamical characteristic of a structured optical field is its helicity [12,15,17] whose density in a medium with negligible dispersion [30] does not differ from the standard definition

$$h = \frac{g}{\omega} \text{Im}(\mathbf{E} \cdot \mathbf{H}^*). \tag{21}$$

It plays a significant role in the field-matter interactions with violation of the rotational symmetry [38–42]. As a scalar characteristic, it holds some features of the energy density and can be considered as its formal analogue for the chiral interactions [43]. The energy and helicity distributions obey the useful inequality setting restriction for the "normalized helicity" $\bar{h}$:

$$|\bar{h}| = \omega\sqrt{\varepsilon\mu}\left|\frac{h}{w}\right| = \frac{2\left|\text{Im}\left(\sqrt{\varepsilon}\mathbf{E} \cdot \sqrt{\mu}\mathbf{H}^*\right)\right|}{\left|\sqrt{\varepsilon}\mathbf{E}\right|^2 + \left|\sqrt{\mu}\mathbf{H}\right|^2} \leq 1 \tag{22}$$

in which the upper limit is reached if $\mathbf{H} = \pm i\sqrt{\varepsilon/\mu}\,\mathbf{E}$, which is realized, for example, in a circularly polarized plane wave (cf. equation (1)).

Before proceeding further, note that, contrary to the total field momentum (13) and the field helicity (21), where the electric and magnetic fields are entangled, the field energy (12), the spin (15) and orbital (16) momenta as well as the spin density (17) can be considered as sums of distinctly separable electric and magnetic contributions expressed by the first and second summands of the corresponding expressions. This division is not formal and reflects specific features of the electromagnetic field interaction with material objects which usually exhibit preferential sensitivity to electric or magnetic field [7,18,31,44]. Consistent description of the electric and magnetic contributions to the field dynamical characteristics, although in the Abraham definitions, has been performed elsewhere: see [34] and the Supplementary Note 2 of [18] for the EW and the Supplemental Material of [7] for the two-wave interference field, which is the prototype of the field in medium 1 of the present model. Therefore, in this paper we will not specially highlight the electric and magnetic contributions but concentrate on the interrelations between the field characteristics in the EW and TR regions.

In the next sections, we present explicit expressions for the field dynamical characteristics both in the EW and TR regions, obtained on the base of their general definitions listed above. In the medium 1 ($-\infty < z < 0$), where the field is formed by superposition of two plane waves (1) and (2), we have to substitute $\mathbf{E} = \mathbf{E}_0 + \mathbf{E}_r$, $\mathbf{H} = \mathbf{H}_0 + \mathbf{H}_r$ described by (1) – (4); in the medium 2, the quantities to be substituted are $\mathbf{E} = \mathbf{E}_t$, $\mathbf{H} = \mathbf{H}_t$ (6). In the following formulas, all the dynamical characteristics should contain, as a common multiplier, the incident wave intensity

$$I_0 = g\left(|E_{0\|}|^2 + |E_{0\perp}|^2\right) \tag{23}$$

which is omitted for the sake of simplicity.

## 3.2. Scalar dynamical characteristics: energy and helicity near the TR interface

We start the calculation with the energy density (12), and after evident transformations, we arrive at the energy distributions in the medium 1 ($-\infty < z < 0$)

$$w^- = 2\varepsilon_1 \left[1 + \sin^2\theta_1 \operatorname{Re}\left(U^- e^{2ik_1 z \cos\theta_1}\right)\right] \tag{24}$$

and medium 2 ($0 < z < \infty$):

$$w^+ = \varepsilon_2 \cosh^2\alpha\, U^+ e^{-2k_2 z \sinh\alpha} \tag{25}$$

where

$$U^- = \frac{1+\tau_0}{2} R_\|^* + \frac{1-\tau_0}{2} R_\perp^*, \quad U^+ = \frac{1-\tau_0}{2}|T_\perp|^2 + \frac{1+\tau_0}{2}|T_\||^2 \tag{26}$$

(superscripts '+' and '–' denote the positive $z > 0$ and negative $z < 0$ half-spaces). The energy distribution (24), (25) is discontinuous at the interface, and, with the help of formula (A1) (Appendix A), the step value can be found in the form

$$\Delta w = \left(w^+ - w^-\right)_{z=0} = \varepsilon_1 \left\{ \frac{\mu_1}{\mu_2} \sin^2\theta_1 \left[\left(1 - \frac{\mu_2}{\mu_1}\right)\frac{1-\tau_0}{2}|T_\perp|^2 + \left(1 - \frac{\varepsilon_2}{\varepsilon_1}\right)\frac{1+\tau_0}{2}|T_\||^2\right] - 2\cos^2\theta_1 \right\}. \tag{27}$$

As could be expected, the energy distribution is fully determined by the Stokes parameter $\tau_0$ (11) which characterizes the relation between the *s*- and *p*-polarized components.

Quite similarly, based on (21), one can find another scalar parameter of the field – its helicity whose spatial density appears in the form

$$h^- = -\frac{1}{\omega}\sqrt{\frac{\varepsilon_1}{\mu_1}} \operatorname{Im}\left\{(\chi_0 - i\sigma_0)\left[1 + R_\perp^* R_\| + \sin^2\theta_1\left(R_\perp^* e^{2ik_1 z\cos\theta_1} + R_\| e^{-2ik_1 z\cos\theta_1}\right)\right]\right\}, \tag{28}$$

$$h^+ = -\frac{1}{\omega}\sqrt{\frac{\varepsilon_2}{\mu_2}} \cosh^2\alpha\, \operatorname{Im}\left[(\chi_0 - i\sigma_0)T_\| T_\perp^*\right] e^{-2k_2 z \sinh\alpha}. \tag{29}$$

In contrast to the energy density, the helicity is determined by the Stokes parameters $\chi_0$ and $\sigma_0$ (11). The role of $\sigma_0$ is intuitively evident, as it expresses the incident wave ellipticity; likewise, the parameter $\chi_0$ may contribute to the reflected wave ellipticity via the phase difference between the complex coefficients $R_\|$ and $R_\perp$. Additional ellipticity in the TR region emerges due to the *z*-dependent phase interrelations between the incident and reflected waves expressed by the exponential terms in (28). These lead to the effects that, in the TR half-space, the polarization ellipticity distribution "generated" by an incident wave with $\chi_0 = 1$ is physically similar to that generated by an incident wave with $\sigma_0 = 1$ but with some spatial shift (see also (A4) in Appendix B; analogous behavior is generally typical in the two-wave interference fields [7]).

Like the energy, the field helicity distribution also exhibits a discontinuity; its step at the interface can be readily found from (28), (29) with account for (5) and (A1):

$$\Delta h = \left(h^+ - h^-\right)_{z=0} = -\frac{1}{2\omega}\sqrt{\frac{\varepsilon_2}{\mu_2}}\sin^2\theta_1\left(\frac{n_1^2}{n_2^2}-1\right)\mathrm{Im}\left[(\chi_0 - i\sigma_0)T_\parallel T_\perp^*\right]. \tag{30}$$

This behavior is illustrated by the numerical example in figure 2. In the half-space $z < 0$, a periodic standing-wave distribution is formed whereas in the EW region all the characteristics decay exponentially. In agreement with (22), normalized helicity never exceeds the unity. Certain local enhancement of the helicity in the EW compared to the near-surface helicity in the medium 1 (25%) is not so impressive for the normalized helicity (4%). At first glance, the "synchronous" variations of $w(z)$ and $h(z)$ in the medium 1 look striking because corresponding equations (24) and (28) show no visible similarity but this is quite natural for $|R_\perp| = |R_\parallel| = 1$ in cases of $\sigma_0 = 1$ and $\chi_0 = 1$ (see relations (A2) and (A3) in the Appendix B). For the *s*-polarization ($\tau_0 = -1$) and *p*-polarization ($\tau_0 = 1$) of the incident wave, when the helicity vanishes identically in the whole space, the energy standing-wave crests and troughs are slightly shifted, while the corresponding changes in the EW region are hardly discernible (thin red curves in figure 2).

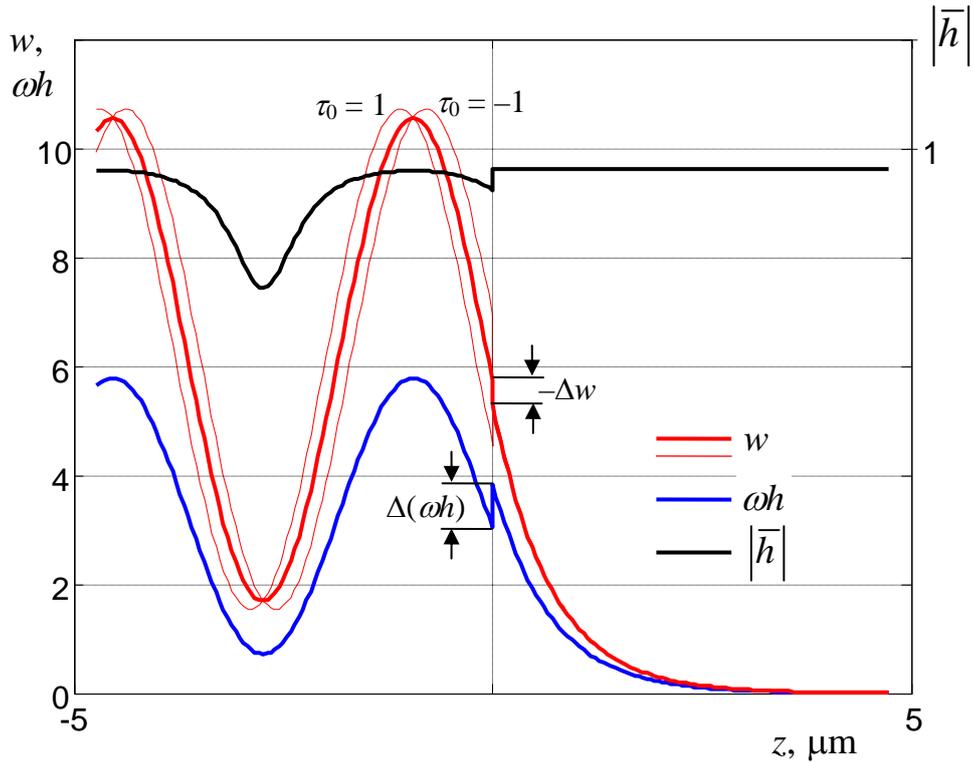

**Figure 2**. Spatial distribution of (red) energy (24), (25), (blue) helicity (28), (29) and (black, right scale) normalized helicity (22) near the interface between heavy flint glass ($z < 0$, $\varepsilon_1 = 3.06$, $\mu_1 = 1$) and water ($z > 0$, $\varepsilon_2 = 1.77$, $\mu_2 = 1$) for the incident circularly polarized ($\sigma_0 = 1$) wave with vacuum wavelength 0.63 μm, $\theta_1 = 60°$ ($\cosh\alpha = 1.14$, $\sinh\alpha = 0.546$). Thin red lines show the energy distribution when the incident wave is linearly polarized with $\tau_0 = \pm 1$.

### 3.3. Field momentum and spin

1) Operating as before, from (13) we obtain the momentum density components

$$p_x^- = \frac{2}{c}\varepsilon_1 n_1 \sin\theta_1 \left[1 + \mathrm{Re}\left(U^- e^{2ik_1 z \cos\theta_1}\right)\right], \tag{31}$$

$$p_x^+ = \frac{1}{c} n_2 \varepsilon_2 \cosh\alpha\, U^+ e^{-2k_2 z \sinh\alpha}; \tag{32}$$

$$p_y^- = -\frac{1}{2c}\varepsilon_1 n_1 \sin 2\theta_1 \,\mathrm{Re}\left[(\chi_0 - i\sigma_0)\left(R_\perp^* e^{2ik_1 z \cos\theta_1} - R_\| e^{-2ik_1 z \cos\theta_1}\right)\right], \tag{33}$$

$$p_y^+ = \frac{1}{c} n_2 \varepsilon_2 \sinh\alpha \cosh\alpha\, \mathrm{Im}\left[(\chi_0 - i\sigma_0) T_\| T_\perp^*\right] e^{-2k_2 z \sinh\alpha}; \tag{34}$$

$$p_z \equiv 0 \quad (-\infty < z < \infty). \tag{35}$$

These results (more exactly, their Abraham analogs) are well known (see, e.g., Refs. [45,46] where they were used for explanation of the Goos–Hänchen and Imbert–Fedorov beam shifts); however, the corresponding energy flow pattern was considered in detail only recently [34]. Equations (31) and (32) describe the usual momentum directed along the main propagation axis $x$. Note the presence of the "extraordinary" [7,18] transverse momentum $p_y$ (33), (34) associated with the Stokes parameters $\chi_0$ and $\sigma_0$ which are also responsible for the field helicity (cf. equations (28) and (29)). The zero value of the normal momentum component $p_z$ (35) is a consequence of the special TR conditions: the whole energy current approaching the interface $z = 0$ is exactly balanced by the reflected current, and the EW does not transport energy away from the interface.

Functions $p_x(z)$ and $p_y(z)$ characterizing the field momentum parallel to the interface, experience simple step discontinuities at $z = 0$ (see figure 3). By using relations (5), (A1) and (31) – (34), these steps can be represented in a compact physically transparent form:

$$\Delta p_x = \left(p_x^+ - p_x^-\right)_{z=0} = \frac{1}{c}\varepsilon_2 n_2 \cosh\alpha \left[\frac{1-\tau_0}{2}|T_\perp|^2\left(1-\frac{\varepsilon_1}{\varepsilon_2}\right) + \frac{1+\tau_0}{2}|T_\||^2\left(1-\frac{\mu_1}{\mu_2}\right)\right], \tag{36}$$

$$\Delta p_y = \left(p_y^+ - p_y^-\right)_{z=0} = -\frac{1}{2c}\varepsilon_2 n_2 \sinh\alpha \cosh\alpha\, \mathrm{Im}\left[(\chi_0 - i\sigma_0) T_\perp^* T_\|\right]\left(\frac{\mu_1}{\mu_2} + \frac{\varepsilon_1}{\varepsilon_2} - 2\right). \tag{37}$$

In the step expressions (36) and (37), the summands proportional to $1 - \mu_1/\mu_2$ ('magnetic') and those proportional to $1 - \varepsilon_1/\varepsilon_2$ ('electric') can be easily singled out, which supports the separate importance of the electric and magnetic contributions discussed above in connection to the dynamical characteristics' definitions (15) – (17).

2) Our next point of interest is the Minkowski spin density (17) that, under the discussed dispersion-free conditions, not only characterizes the intrinsic angular momentum of the field but also "generates" the spin linear momentum (15):

$$s_x^- = -\frac{1}{\omega}\varepsilon_1 \sin\theta_1 \,\mathrm{Im}\left[(\chi_0 - i\sigma_0)\left(1 + R_\perp^* e^{2ik_1 z \cos\theta_1} + R_\| e^{-2ik_1 z \cos\theta_1} + R_\perp^* R_\|\right)\right], \tag{38}$$

$$s_x^+ = -\frac{1}{\omega}\varepsilon_2 \cosh\alpha\, \mathrm{Im}\left[(\chi_0 - i\sigma_0) T_\| T_\perp^*\right] e^{-2k_2 z \sinh\alpha}; \tag{39}$$

$$s_y^- = -\frac{2}{\omega}\varepsilon_1 \sin\theta_1 \cos\theta_1 \,\mathrm{Im}\left(U^- e^{2ik_1 z \cos\theta_1}\right), \tag{40}$$

$$s_y^+ = -\frac{1}{\omega}\varepsilon_2 \sinh\alpha \cosh\alpha\, U^+ e^{-2k_2 z \sinh\alpha}; \tag{41}$$

$$s_z^- = -\frac{1}{\omega}\varepsilon_1 \cos\theta_1 \,\mathrm{Im}\left[(\chi_0 - i\sigma_0)\left(1 - R_\perp^* R_\|\right)\right], \tag{42}$$

$$s_z^+ = 0. \tag{43}$$

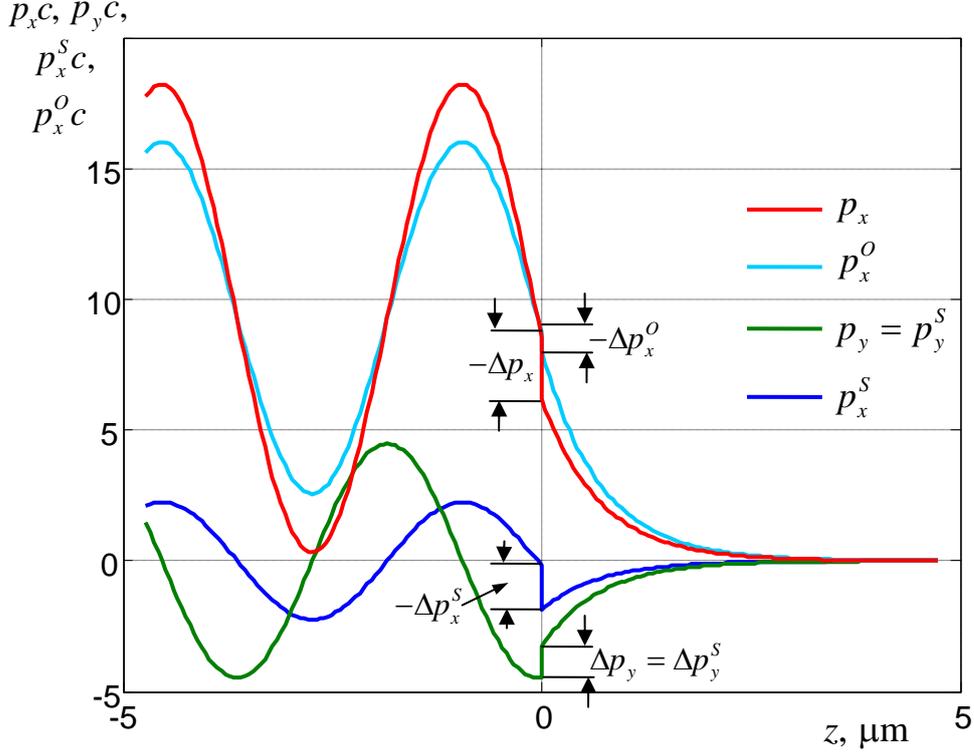

**Figure 3**. Spatial distribution of (red) longitudinal component $p_x$ of the momentum (31), (32), (green) transverse "extraordinary" momentum $p_y$ (33), (34) that coincides with the transverse spin momentum $p_y^S$ (47), (48), (blue) longitudinal spin momentum $p_x^S$ (45), (46) and (light blue) longitudinal orbital momentum $p_x^O$ (49), (50); the field configuration is the same as in figure 2.

Here the $x$-component (38), (39) (see figure 4, red curve) is the most familiar and describes the "ordinary" spin oriented parallel to the propagation direction and owing to the ellipticity of the wave polarization in the transverse $yz$ plane. Comparison with (28) and (29) shows that it is connected to the field helicity which, in turn, is associated with the ellipticity of the incident, reflected or refracted wave. In particular, the contribution $s_x^+$ (39) appears due to ellipticity of the EW polarization expressed by non-zero value of $-\operatorname{Im}\left[\left(\chi_0 - i\sigma_0\right)T_\| T_\perp^*\right] \propto \sigma_t$.

Vanishing $s_z^+$ in the EW is quite expected from the symmetry but, in fact, it is formed by opposite non-zero electric and magnetic contributions which exactly cancel each other [18] (separate existence of these contributions can be revealed through the mechanical torque exerted on a dual asymmetric material object). In contrast, non-zero $s_z^-$ in the medium 1 (black line in figure 4) results from the geometric summation of spins carried by the incident and reflected waves [7,9] and owes to the difference in arguments of the unimodular complex quantities $R_\|$ and $R_\perp$. Had the reflection been perfect (that is, $R_\| = R_\perp$ and the only change in the polarization state had been the ellipticity sign reversal, $\sigma_r = -\sigma_0$), the normal spin component would have vanished, $s_z^- = 0$.

The most interesting result is expressed by formulae (40) and (41) which describe the specific transverse spin (green line in figure 4) that is of a particular interest in structured optical fields [7,9,10,14,15,17,18,31,46,47]; its general and comprehensive description has been recently developed [23]. In addition to the "anomalous" direction (orthogonal to all the wavevectors of the incident, reflected and evanescent (9) waves), its remarkable feature is that it arises independently of the incident wave state of polarization and does not require any initial ellipticity.

In contrast to the energy, helicity and momentum distributions, considered above, the spin components parallel to the interface are continuous: $\Delta s_x = \Delta s_y = 0$ (see figure 4). This feature is related with the choice of Minkowski spin as the true intrinsic angular momentum [31] of the field in non-dispersive dielectric media and sharply contradicts to the earlier analyses dealing with the Abraham definitions [17,34].

### 3.4. Spin – orbital momentum decomposition

At this point, we focus on the spin (15) and orbital (16) momenta's distributions. Knowledge of the spin components (38) – (41) and equation (15) suggest a direct way to render the momentum decomposition (14) via the simple prescription

$$\mathbf{p}^S = \frac{1}{2}\begin{pmatrix} -\partial s_y/\partial z \\ \partial s_x/\partial z \\ 0 \end{pmatrix}, \quad \mathbf{p}^O = \mathbf{p} - \mathbf{p}^S. \tag{44}$$

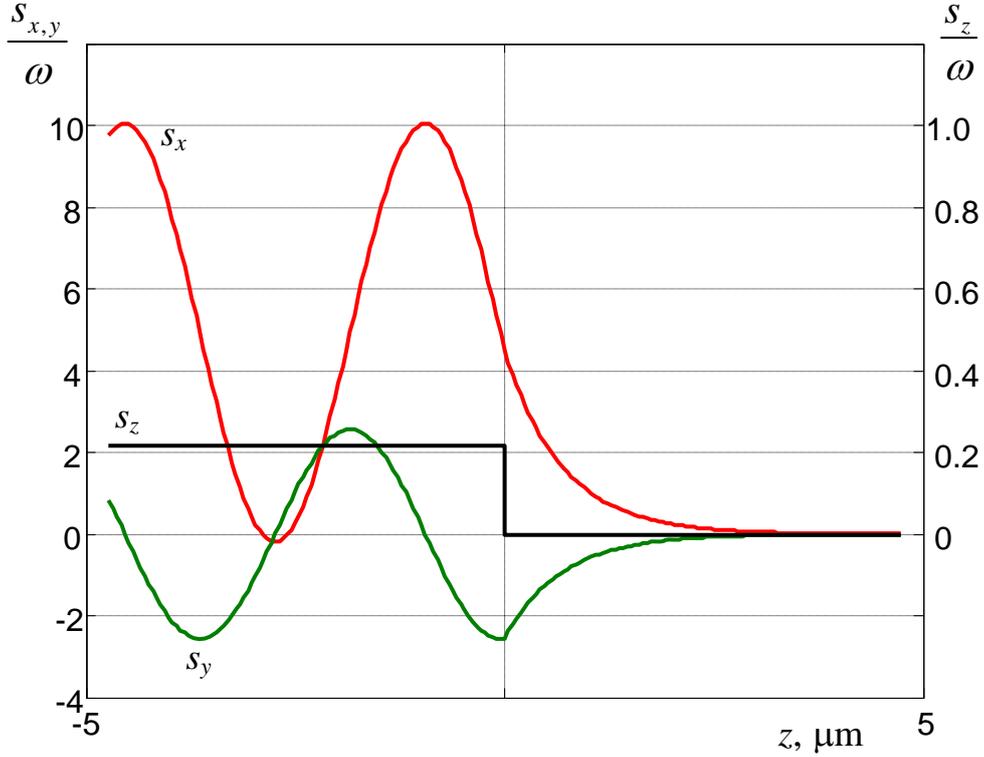

**Figure 4**. Spatial distribution of (red) longitudinal component of spin $s_x$ (38), (39), (green) transverse "extraordinary" spin $s_y$ (40), (41), (black, right scale) spin $s_z$ normal to the interface (42); the field configuration is the same as in figure 2.

By substituting (38) – (41) into (44), we obtain the following expressions for the spin momentum density contributions:

$$p_x^{S-} = 2\frac{n_1}{c}\varepsilon_1 \sin\theta_1 \cos^2\theta_1 \operatorname{Re}\left(U^- e^{2ik_1 z\cos\theta_1}\right), \tag{45}$$

$$p_x^{S+} = -\frac{n_2}{c}\varepsilon_2 \sinh^2\alpha \cosh\alpha \, U^+ e^{-2k_2 z \sinh\alpha} ; \quad (46)$$

$$p_y^{S-} = -\frac{n_1}{c}\varepsilon_1 \sin\theta_1 \cos\theta_1 \, \text{Re}\left[(\chi_0 - i\sigma_0)\left(R_\perp^* e^{2ik_1 z \cos\theta_1} - R_\| e^{-2ik_1 z \cos\theta_1}\right)\right], \quad (47)$$

$$p_y^{S+} = \frac{n_2}{c}\varepsilon_2 \sinh\alpha \cosh\alpha \, \text{Im}\left[(\chi_0 - i\sigma_0)T_\| T_\perp^*\right] e^{-2k_2 z \sinh\alpha} . \quad (48)$$

Comparison of equations (47), (48) and (33), (34) shows that $p_y^S = p_y$, and, consequently, $p_y^O = 0$: the transverse momentum is completely of the spin nature, and all its properties can be immediately obtained from (33), (34) and (37). The $z$-component of the spin momentum vanishes due to first equation (44), and in combination with (35) it means that $p_z^O = 0$ in the entire space. Therefore, the whole orbital (canonical) momentum is $x$-oriented and, according to (31), (32), (45), (46) and second equation (44), equals to

$$p_x^{O-} = \frac{2}{c}\varepsilon_1 n_1 \sin\theta_1 \left[1 + \sin^2\theta_1 \, \text{Re}\left(U^- e^{2ik_1 z \cos\theta_1}\right)\right], \quad (49)$$

$$p_x^{O+} = \frac{n_2}{c}\varepsilon_2 \cosh^3\alpha \, U^+ e^{-2k_2 z \sinh\alpha} \quad (50)$$

(see figure 3, light blue curve). With account for the first equation (5) and relations (24), (25) this entails that in the whole space

$$p_x^O = \frac{n_1}{c}\sin\theta_1 \cdot w = \frac{n_2}{c}\cosh\alpha \cdot w : \quad (51)$$

the orbital momentum is distributed proportionally to the energy, which is typical for other fields formed by EW and plane waves [7,31]. Accordingly, the characteristic features of its distribution can be easily derived from (24) – (27). In particular, the spin and orbital longitudinal momenta $p_x^S$ and $p_x^O$ are discontinuous at the interface (see figure 3); the orbital momentum step at $z = 0$ equals

$$\Delta p_x^O = \left(p_x^{O+} - p_x^{O-}\right)_{z=0} = \frac{n_1}{c}\sin\theta_1 \Delta w \quad (52)$$

(cf. equation (27)). The corresponding spin momentum step can be derived from (52), (27), (36) and second equation (44) in the form

$$\Delta p_x^S = \left(p_x^{S+} - p_x^{S-}\right)_{z=0} = \frac{n_1}{c}\sin\theta_1 \left[-\frac{1+\tau_0}{2}|T_\||^2 \varepsilon_2\left(\sinh^2\alpha + \frac{\mu_1}{\mu_2}\cos^2\theta_1\right)\right.$$
$$\left. -\frac{1-\tau_0}{2}|T_\perp|^2 \varepsilon_2\left(\sinh^2\alpha + \frac{\varepsilon_1}{\varepsilon_2}\cos^2\theta_1\right) + 2\varepsilon_1 \cos^2\theta_1\right]. \quad (53)$$

Several important conclusions can be inferred immediately from the spin and orbital momentum distributions (45) – (53) (see also the illustrations in figures 3 and 4). First to note, since the spin components $s_x$ and $s_y$ (cf. equations (38) – (41) and the red and green curves in figure 4) are continuous, the differentiation in (44) produces no singularities in the spin momentum distributions that were predicted for the surface-polariton and the TR fields earlier [17,34]. Accordingly, the "volume" components (45) – (48) provide exhaustive spin momentum characterization. The special "boundary" spin and orbital flows proportional to the Dirac delta-function $\delta(z)$ appeared in [17,34] because the dynamical characteristics were grounded on the Abraham momentum; the suggested singular terms were hardly interpretable but played the principal role ensuring that the total spin momentum of the field vanishes, which is required by the general spin momentum theory [3,8]. Now one can easily persuade that the present forms of the "volume" spin momentum distributions (45) – (48) satisfy this crucial requirement automatically (see Appendix C, formulae (A6) – (A10)).

Next, relation (51) complies with the orbital momentum interpretation as the canonical momentum responsible for the field energy transport [18,37]. The corresponding energy flow density (18) is expectedly *x*-directed and is described by different equations in the lower and upper half-spaces,

$$q_x^- = \frac{c \sin \theta_1}{n_1} w^-, \quad q_x^+ = \frac{c \cosh \alpha}{n_2} w^+ = \frac{c n_1 \sin \theta_1}{n_2^2} w^+. \tag{54}$$

This quantity is also discontinuous with the boundary step

$$\Delta q_x = \left(q_x^+ - q_x^-\right)_{z=0} = c n_1 \sin \theta_1 \left[ \frac{1+\tau_0}{2} \frac{\sin^2 \theta_1}{\mu_2} |T_\parallel|^2 \left(\frac{n_1^2}{n_2^2} - \frac{\varepsilon_2}{\varepsilon_1}\right) \right.$$
$$\left. + \frac{1-\tau_0}{2} \frac{\sin^2 \theta_1}{\mu_2} |T_\perp|^2 \left(\frac{n_1^2}{n_2^2} - \frac{\mu_2}{\mu_1}\right) - \frac{2}{\mu_1} \cos^2 \theta_1 \right]. \tag{55}$$

But more noteworthy is that equations (54) determine different local group velocities $v_g = q_x/w$ of the energy transportation in each medium

$$v_g^- = \frac{c \sin \theta_1}{n_1}, \quad v_g^+ = \frac{c \cosh \alpha}{n_2}, \tag{56}$$

and $v_g^+$ can be superluminal [17,18], in contrast to the similar ratio for the total momentum (32) $\left(c^2/\varepsilon_2 \mu_2\right)\left(p_x^+/w^+\right) = c/(n_2 \cosh \alpha)$. This is accompanied by the continuous subluminal *x*-directed phase velocity $v_p = c/(n_1 \sin \theta_1) = c/(n_2 \cosh \alpha)$, and, remarkably, the typical waveguide relations [1]

$$v_g^- v_p = \frac{c^2}{n_1^2}, \quad v_g^+ v_p = \frac{c^2}{n_2^2} \tag{57}$$

hold for the waves running in the *x*-direction within both media.

The superluminal local group velocity in the EW region does not imply superluminal energy propagation: analysis of the quantities $w$ and $q_x$ integrated over the whole field shows that any wave packet moves with velocity less than *c* [31,32]. However, the possibility of $v_g^+ > c$ admitted by the second equation (56) is coupled with the anomalously high orbital momentum $p_x^{O+} = \cosh^2 \alpha \cdot p_x^+$ that exceeds the momentum of a plane wave with the same local intensity [17,18]. This "super-momentum" can be detected, e.g., via the force exerted on a small particle immersed in the EW field, which proportionally exceeds the force from a plane wave (this was predicted in Ref. [48] analytically for a dipole particle and can be supported by numerical calculations for spherical Mie particles with larger size). Also, it can be measured indirectly, via the resonant frequency of the moving atom – EW interaction which experiences the Doppler shift proportional to the momentum exchanged between the field and the atom [49,50].

General features of the spin and orbital momentum distributions are illustrated by figure 3. One can see the obvious similarity between $p_x^O$ (light blue curve in figure 3) and $w$ (red curve in figure 2) distributions; in full agreement to the forms of equations (31) and (32), (45) and (46), (49) and (50), the variable parts of $p_x$ (red curve), $p_x^O$ (light blue) and $p_x^S$ (blue) are mutually proportional and reach maxima and minima simultaneously and in accord with the standing-wave energy distribution (figure 2). The spin momentum density $p_y^S$ (green curve in figure 3), oppositely, shows the quarter-period spatial shift because it is proportional to the energy gradient. In more detail, the interplay of different dynamical characteristics in the TR region is commented in Appendix B, relations (A2) – (A5).

## 4. Conclusion

In a short summary we first of all underline that this paper presents a full collection of explicit expressions describing the spatial distributions of the main dynamical characteristics: energy, momentum, spin and helicity of the structured electromagnetic field formed in the process of total reflection. In addition to other known researches, we consistently analyze the distributions on both sides of the total-reflection interface, in the evanescent-wave as well as in the total-reflection regions, and pay a special attention to the characteristics' behavior upon crossing the boundary. Since many of the dynamical characteristics analyzed in the paper play important role in various light–matter interactions [6,7,10,12–14,18,43], main results of the paper will be useful in practical problems of engineering the structured optical fields and their employment for microscopy [11], optically-controlled microdevices [10,13,14], information processing, sorting, dispatching and precise manipulation of micro- and nano-objects [12,23,43].

At the same time, our findings offer a certain fundamental aspect. Actually, here we present, together with references [31–33], one of the first examples of consistent decomposition of the Poynting momentum into the orbital (canonical) and spin parts in an inhomogeneous medium, and the peculiar properties of these momentum constituents are considered in many detail. Although in the dispersion-free approximation (but, in contrast to the surface polariton considered in [31,32], for the essentially 3D geometry), this procedure could be completed only due to relying on the recent deep investigation of the principal features of electromagnetic field in material media [28–31]. For the correct specification of the spin and orbital momenta, a proper definition of the spin density is of crucial importance, and the results of [28–31] enabled us to substantiate that it is the Minkowski spin that is preferable, at least in the problems relating a sharp interface and the total-reflection conditions. The medium inhomogeniety provides an additional support to the Minkowski form of the spin density in the material media: the fact is that its components parallel to the interface are the only dynamical characteristics that are continuous at the boundary. This is methodologically valuable as it results in the non-singular and physically meaningful distributions of the spin and orbital momenta in the whole space and thus enables the physically consistent description of their mechanical action [7,12,18,44].

This observation looks hopeful for possible generalizations of the approach used in this paper to more complex inhomogeneous media. As a first step, efforts to study the spatial distribution of the dynamical characteristics in stratified and one-dimensional layered media may be prospective. The generalization to dispersive media looks straightforward from the mathematical point of view: it only requires replacement of the media's permittivity and permeability in (12) and (15) – (17) by the dispersion-modified quantities, following to the prescriptions of [31–33]. However, then the dielectric and magnetic constants in the expressions for the dynamical characteristics (12), (15) – (17) will differ from the 'simple' permittivities and permeabilities that determine the reflected (2) and transmitted (6) fields via the corresponding coefficients (4), (8). As a result, the description of the dynamical characteristics will be more complicated and less transparent physically, but one can expect many new interesting details in their structure and spatial distributions.

## Appendix A: Relations between the transmission and reflection coefficients

Equations (4) and (8) together with the definition of the parameter $\alpha$ (5) entail a number of useful relations that are repeatedly used in the algebraic transformations throughout the paper:

$$1 + R_\perp = T_\perp, \quad 1 + R_\parallel = \sqrt{\frac{\varepsilon_2 \mu_1}{\varepsilon_1 \mu_2}} T_\parallel,$$

$$|T_\parallel|^2 = 2\frac{\cos\theta_1}{\sinh\alpha}\sqrt{\frac{\varepsilon_1\mu_2}{\varepsilon_2\mu_1}}\,\mathrm{Im}\,R_\parallel^* = 2\frac{\varepsilon_1\mu_2}{\varepsilon_2\mu_1}\mathrm{Re}(1+R_\parallel), \quad |T_\perp|^2 = 2\frac{\cos\theta_1}{\sinh\alpha}\sqrt{\frac{\varepsilon_1\mu_2}{\varepsilon_2\mu_1}}\,\mathrm{Im}\,R_\perp^* = 2\,\mathrm{Re}(1+R_\perp),$$

$$R_\perp^* - R_\parallel = \frac{i}{2}\frac{\sinh\alpha}{\cos\theta_1}\left(1+\frac{\varepsilon_2\mu_1}{\varepsilon_1\mu_2}\right)T_\parallel T_\perp^*, \quad R_\perp^* + R_\parallel = \frac{1}{2}\sqrt{\frac{\varepsilon_2\mu_1}{\varepsilon_1\mu_2}}\left(1-\frac{\sinh^2\alpha}{\cos^2\theta_1}\right)T_\parallel T_\perp^*. \tag{A1}$$

## Appendix B. Details of the standing-wave pattern in the TR region

Behavior of the field characteristics in medium 1 can be better understood when taking into account that the reflection coefficients possess unit modules and can be represented in forms

$$R_\perp = e^{i\delta_\perp}, \quad R_\parallel = e^{i\delta_\parallel}.$$

Then the real and imaginary parts of the factor $U^- \exp(2ik_1 z\cos\theta_1)$ that determine spatial distributions in equations (24), (31), (40), (45), (50) can be written as

$$\mathrm{Re}\left(U^- e^{2ik_1 z\cos\theta_1}\right) = \cos\left(2k_1 z\cos\theta - \frac{\delta_\perp + \delta_\parallel}{2}\right)\cos\frac{\delta_\perp - \delta_\parallel}{2}$$
$$+\tau_0 \sin\left(2k_1 z\cos\theta - \frac{\delta_\perp + \delta_\parallel}{2}\right)\sin\frac{\delta_\parallel - \delta_\perp}{2}, \tag{A2}$$

$$\mathrm{Im}\left(U^- e^{2ik_1 z\cos\theta_1}\right) = \sin\left(2k_1 z\cos\theta - \frac{\delta_\perp + \delta_\parallel}{2}\right)\cos\frac{\delta_\perp - \delta_\parallel}{2}$$
$$-\tau_0 \cos\left(2k_1 z\cos\theta - \frac{\delta_\perp + \delta_\parallel}{2}\right)\sin\frac{\delta_\parallel - \delta_\perp}{2}. \tag{A3}$$

The variable part of the helicity distribution (28) is proportional to the expression

$$\mathrm{Im}\left[(\chi_0 - i\sigma_0)\left(R_\perp^* e^{2ik_1 z\cos\theta_1} + R_\parallel e^{-2ik_1 z\cos\theta_1}\right)\right] =$$
$$= \cos\left(2k_1 z\cos\theta - \frac{\delta_\perp + \delta_\parallel}{2}\right)\left(\chi_0 \sin\frac{\delta_\parallel - \delta_\perp}{2} - \sigma_0 \cos\frac{\delta_\parallel - \delta_\perp}{2}\right); \tag{A4}$$

similarly, the variable part of the transverse momentum distribution (33), (47) behaves as

$$\mathrm{Re}\left[(\chi_0 - i\sigma_0)\left(R_\perp^* e^{2ik_1 z\cos\theta_1} - R_\parallel e^{-2ik_1 z\cos\theta_1}\right)\right] =$$
$$= \sin\left(2k_1 z\cos\theta - \frac{\delta_\perp + \delta_\parallel}{2}\right)\left(-\chi_0 \sin\frac{\delta_\parallel - \delta_\perp}{2} + \sigma_0 \cos\frac{\delta_\parallel - \delta_\perp}{2}\right). \tag{A5}$$

Comparison of (A2) and (A4) explains the concerted variations of the energy and helicity densities in the case $\tau_0 = 0$ (figure 2); slight shifts of the energy distribution for $\tau_0 = \pm 1$ are caused by the second summand of (A2) whose magnitude is limited by the relatively small value of $\sin\left[(\delta_\parallel - \delta_\perp)/2\right]$. Equations (A3) and (A5) being confronted with (A2) elucidate the quarter-period shifts of the transverse spin (40) and the transverse momentum (33), (47) distributions (green curves in figures 3, 4) with respect to the energy and the longitudinal momentum distributions obtained under the condition $\tau_0 = 0$ (figures 2, 3).

## Appendix C. Calculation of the total spin momentum of the field in the entire space

According to (45) – (48), the total spin momentum of the field consists of the $x$- and $y$-components which are proportional to the following integrals taken over the whole $z$-axis:

$$P_{x,y}^S = \int_{-\infty}^{\infty} p_{x,y}^S(z)\,dz = \int_{-\infty}^{0} p_{x,y}^{S-}(z)\,dz + \int_{0}^{\infty} p_{x,y}^{S+}(z)\,dz. \tag{A6}$$

The direct evaluation of this expression is impossible because the first integral in the right-hand side with $p_{x,y}^{S-}(z)$ taken in the form (45) and (47) diverges. To make the integral meaningful, the integrand quantities $p_{x,y}^{S-}(z)$ should be properly regularized. To this end, we imply the infinitesimal absorption in the medium 1 so that expressions (38) and (40) are replaced by

$$s_x^- = \frac{1}{\omega}\varepsilon_1 \sin\theta_1 \, \text{Im}\left[(\chi_0 - i\sigma_0)\left(1 + R_\perp^* e^{2ik_1 z\cos\theta_1} + R_\| e^{-2ik_1 z\cos\theta_1} + R_\perp^* R_\|\right)\right]e^{\kappa z}, \quad \kappa \to 0. \tag{A7}$$

$$s_y^- = -\frac{2}{\omega}\varepsilon_1 \sin\theta_1 \cos\theta_1 \, \text{Im}\left(U^- e^{2ik_1 z\cos\theta_1}\right)e^{\kappa z}, \quad \kappa \to 0; \tag{A8}$$

Substituting this into the first equation (44) one finds, instead of (45) and (47),

$$p_y^{S-} = \frac{n_1}{2c}\varepsilon_1 \sin\theta_1 \, \text{Im}\left\{(\chi_0 - i\sigma_0)\left[\frac{\kappa}{k_1}(1 + R_\perp^* R_\|) + \left(2i\cos\theta_1 + \frac{\kappa}{k_1}\right)R_\perp^* e^{2ik_1 z\cos\theta_1}\right.\right.$$

$$\left.\left. -\left(2i\cos\theta_1 - \frac{\kappa}{k_1}\right)R_\| e^{-2ik_1 z\cos\theta_1}\right]\right\}e^{\kappa z}, \quad \kappa \to 0; \tag{A9}$$

$$p_x^{S-} = \frac{n_1}{c}\varepsilon_1 \sin\theta_1 \cos\theta_1 \, \text{Im}\left[\left(2i\cos\theta_1 + \frac{\kappa}{k_1}\right)U^- e^{2ik_1 z\cos\theta_1}\right]e^{\kappa z}, \quad \kappa \to 0. \tag{A10}$$

Now the integrals over the lower half space $(-\infty, 0)$ can be easily evaluated; the rest of the integrals in (A6) cause no problems because the decaying exponential factor contained in $p_{x,y}^{S+}$ (46), (48) ensures their convergence. After the integrations are performed, passage to the limit $\kappa \to 0$, with taking (A1) into account, immediately entails that quantity (A6) vanishes and, therefore the total spin momentum of the field is zero.

**Appendix D: Distribution of the imaginary Poynting momentum**

The imaginary Poynting (reactive) momentum appears in the description of a monochromatic optical field where the time-averaged Poynting vector is determined via the real part of the complex vector product $\mathbf{E}^* \times \mathbf{H}$ [1]. Its imaginary part is thus associated with the "reactive or stored energy and its alternating flow" [1]. The dynamical meaning of the corresponding imaginary counterpart of the real Poynting momentum is not well established; however, recent research has demonstrated that it can be a source of the mechanical action on a magnetodielectric dipole particle [7,17,24]. For this reason, it can be considered in the common context with the traditional dynamical characteristics of electromagnetic field analyzed in this paper.

In the spirit of the Minkowski paradigm expressed by the momentum definition (13), we define the reactive momentum as

$$\boldsymbol{\pi} = \frac{g}{c}\text{Im}\left(\varepsilon\mu \mathbf{E}^* \times \mathbf{H}\right). \tag{A11}$$

Substituting here expressions (1) – (3) and (6), (7) and operating the same way as on deriving equations (31) – (34), after normalizing by the common multiplier (23) we obtain

$$\pi_x^- = \pi_x^+ = 0; \tag{A12}$$

$$\pi_y^- = \frac{1}{c}\varepsilon_1 n_1 \sin\theta_1 \cos\theta_1 \, \text{Im}\left[(\chi_0 - i\sigma_0)\left(R_\perp^* e^{2ik_1 z\cos\theta_1} - R_\| e^{-2ik_1 z\cos\theta_1}\right)\right], \tag{A13}$$

$$\pi_y^+ = \frac{1}{c}n_2\varepsilon_2 \sinh\alpha \cosh\alpha \, \text{Re}\left[(\chi_0 - i\sigma_0)T_\| T_\perp^*\right]e^{-2k_2 z\sinh\alpha}; \tag{A14}$$

$$\pi_z^- = \frac{2}{c}\varepsilon_1 n_1 \cos\theta_1 \, \text{Im}\left[\left(\frac{1-\tau_0}{2}R_\perp^* - \frac{1+\tau_0}{2}R_\|^*\right)e^{2ik_1 z\cos\theta_1}\right], \tag{A15}$$

$$\pi_z^+ = -\frac{1}{c} n_2 \varepsilon_2 \sinh\alpha \, U^+ e^{-2k_2 z \sinh\alpha}. \tag{A16}$$

The reactive momentum components are discontinuous at the interface with steps

$$\left(\pi_y^+ - \pi_y^-\right)_{z=0} = \frac{1}{2c} n_2 \varepsilon_2 \sinh\alpha \cosh\alpha \, \mathrm{Re}\left[(\chi_0 - i\sigma_0) T_\parallel T_\perp^*\right] \left(2 - \frac{\varepsilon_1}{\varepsilon_2} - \frac{\mu_1}{\mu_2}\right),$$

$$\left(\pi_z^+ - \pi_z^-\right)_{z=0} = \frac{1}{c} n_2 \varepsilon_2 \sinh\alpha \left[\frac{1+\tau_0}{2} |T_\parallel|^2 \left(\frac{n_1^2}{n_2^2} - 1\right) - \frac{1-\tau_0}{2} |T_\perp|^2 \left(\frac{n_1^2}{n_2^2} + 1\right)\right].$$